\documentclass{ws-ijmpd}

\begin{document}

\markboth{D. B. de Freitas et al.}
{Wavelet analysis of stellar rotation and other periodicities}

%
\catchline{}{}{}{}{}
%

\title{WAVELET ANALYSIS OF STELLAR ROTATION AND OTHER PERIODICITIES: FROM THE SUN TO CoRoT}

\author{D. B. de Freitas, I. de C. Le{\~a}o, B. L. Canto Martins and J. R. De Medeiros}

\address{Departamento de F\'{\i}sica, Universidade Federal do Rio Grande do Norte \\ 
Natal, Rio Grande do Norte, 59072-970, Brazil \\
danielbrito@dfte.ufrn.br}

\maketitle

\begin{history}
\received{Day Month Year}
\revised{Day Month Year}
\comby{Managing Editor}
\end{history}

\begin{abstract}
The wavelet transform has been used for numerous studies
in astrophysics, including
signal--noise periodicity and decomposition as well as the
signature of differential rotation in stellar light curves. In
the present work, we apply the
Morlet wavelet with an adjustable parameter $a$, which
can be
fine-tuned to produce optimal resolutions of time and
frequency, and the Haar
wavelet for decomposition at levels of light
curves. We use the
WaveLab--package (library of Matlab routines for wavelet
analysis) for the
decomposition and a modified version of Colorado--package
for the wavelet
maps of synthetic and observed light curve.
From different applications, including Virgo/SoHO, NSO/Kitt Peak, Voyager 1 and Sunspot data
and synthetic light curve produced by different simulators,
we show that this technique is a solid procedure to extract the stellar
rotation period and
possible variations due to
active regions evolution. In this paper we show the Morlet Wavelet Amplitude Maps, respectively corresponding
to oscillations in the photospheric magnetic field of the Sun (NSO/Kitt
Peak data), the daily averages of the magnetic field strength B versus time measured
by Voyager 1 (V1) during 1978, and synthetic light curve produced by A. F.
Lanza. We can also identify the noise
level, as well as the contribution
for the light curves produced by intensity, variability
and mean lifetime of spots.
Thus, we can identify clearly the temporal evolution of
the rotation period in relation to other
periodicity phenomena affecting stellar light curves. In
this context, because the wavelet technique
is a powerful tool to solve, in particular, not trivial
cases of light curves, we are confident that
such a procedure will play an important role on the CoRoT
data analysis.
\end{abstract}


\section{Introduction}	
Signal processing plays a central role of a truly enormous range of the Astrophysical problems. These incluse, for example, rotational periodicity, behavior of stellar magnetic ativity (flares, spots, faculae and plages), oscilations and noise. Much of traditional signal processing has relied upon a relatively small class of problems, for example, stationary and cyclostationary signal characterized by a form of translational invariance. It is not suprising that the Fourier Transform (FT) is considered the key tool in the analysis and manipulation of these problems. But, there are many signals whose defining characteristic is their invariance not to translation but rather to $scale$, i.e., the process exhibit a dependence on different time scales. In addition, while the FT plays a central role in the analysis and manipulation of both statiscally and deterministically translation-invariant signal, the \textit{Wavelet Transform} (WT) plays an analogous role for the kinds of scale-invariant signal.

Wavelet analysis is becoming a common tool for analyzing localized
variations of power within a time series. By decomposing a time series
into time--frequency plane, it is possible to determine both the dominant
modes of variability and how those modes vary in time (see Torrence \&
Compo 1998 for futher details)\cite{toor}. Contrary to classical Fourier analysis
that decomposes a signal into different sines and cosines which are not
bounded in time, the wavelet transform uses functions characterized by
scale (period) and position in time.
The wavelet transform has been used for numerous studies
in astrophysics, including
signal--noise periodicity and decomposition as well as the
signature of differential rotation in stellar light curves. In
the present work, we apply the
Morlet wavelet with an adjustable parameter $a$ (see Section 2), which
can be
fine-tuned to produce optimal resolutions of time and
frequency, where for large values of $a$ give better time resolution.
The use of the Morlet wavelet allow the best trade--off between time and
frequency
resolution, as the gaussian function is its own Fourier transform (Oliver
et al. 1998)\cite{oliver}.
In addition, we apply the Haar
wavelet for decomposition at levels of light
curves. We use the
WaveLab--package (library of Matlab routines for wavelet
analysis) for the
decomposition and a modified version of Colorado--package
for the wavelet
maps of synthetic and observed light curve.
We also apply the wavelet analysis of signals with gaps and of unevenly
spaced data (Frick et al. 1998)\cite{frick}.

\section{Procedures and data analysis}
Many wavelet families can be proposed, depending on the nature of problem. The most commonly used in astrophysical applications is the Morlet Wavelet, that can be defined as being the generalization of the windowed Fourier Transform. The wavelet transform uses a window whose width is a function
of the frequency. Several types of wavelets can be
used. However, if the signal is sinusoidal, the wavelet should
also be chosen to be sinusoidal. On the other hand, the Morlet wavelet (Grossmann
\& Morlet 1984)\cite{gross} represents a sinusoidal oscillation contained within a Gaussian
envelope. Then the wavelet transform can be written as
\begin{equation}
\label{c1}
W(t)=e^{-a[\nu(t-\tau)]^{2}}e^{-i2\pi \nu(t-\tau)},
\end{equation}
represents a sinusoidal oscillation contained within a Gaussian
package. Then the wavelet transform is given by
\begin{equation}
\label{c2}
WT(\nu,\tau)=\sqrt{2\pi \nu}\int S(t)W^{*}[\nu(t-\tau)]dt,
\end{equation}
where $W^{*}$ is the complex conjugate of $W$ and $S(t)$ is the signal. Where an element of the discrete wavelet
transform is
\begin{equation}
\label{c3}
WT(\nu,\tau)_{j}=\sqrt{2\pi \nu}S_{j}(t_{j+1}-t_{j})e^{-a[\nu(t_{j}-\tau)]^{2}}e^{-i2\pi \nu(t_{j}-\tau)}.
\end{equation}

The $a$--parameter controls the resolution in
both frequency and time (Baudin et al. 1994)\cite{baudin}:
\begin{equation}
\label{c4}
\Delta \tau=\frac{1}{\nu}\sqrt{\frac{ln2}{a}},
\end{equation}

\begin{equation}
\label{c5}
\Delta \nu=\frac{\nu}{\pi}\sqrt{aln2},
\end{equation}
and therefore, we have the following uncertainty relationship:
\begin{equation}
\label{c6}
\Delta\tau \cdot\Delta\nu=\frac{ln2}{\pi}.
\end{equation}

The relative frequency resolution is uniquely determined
by $a$--parameter, whereas the time resolution depends on the
frequency itself to hold the number of oscillations inside the
wavelet constant. We choose $a$ = 0.005 in our analysis in order
to balance the time and frequency resolution\cite{hel}.

\begin{figure}[!htb]
\begin{center}
\includegraphics[width=0.85\textwidth]{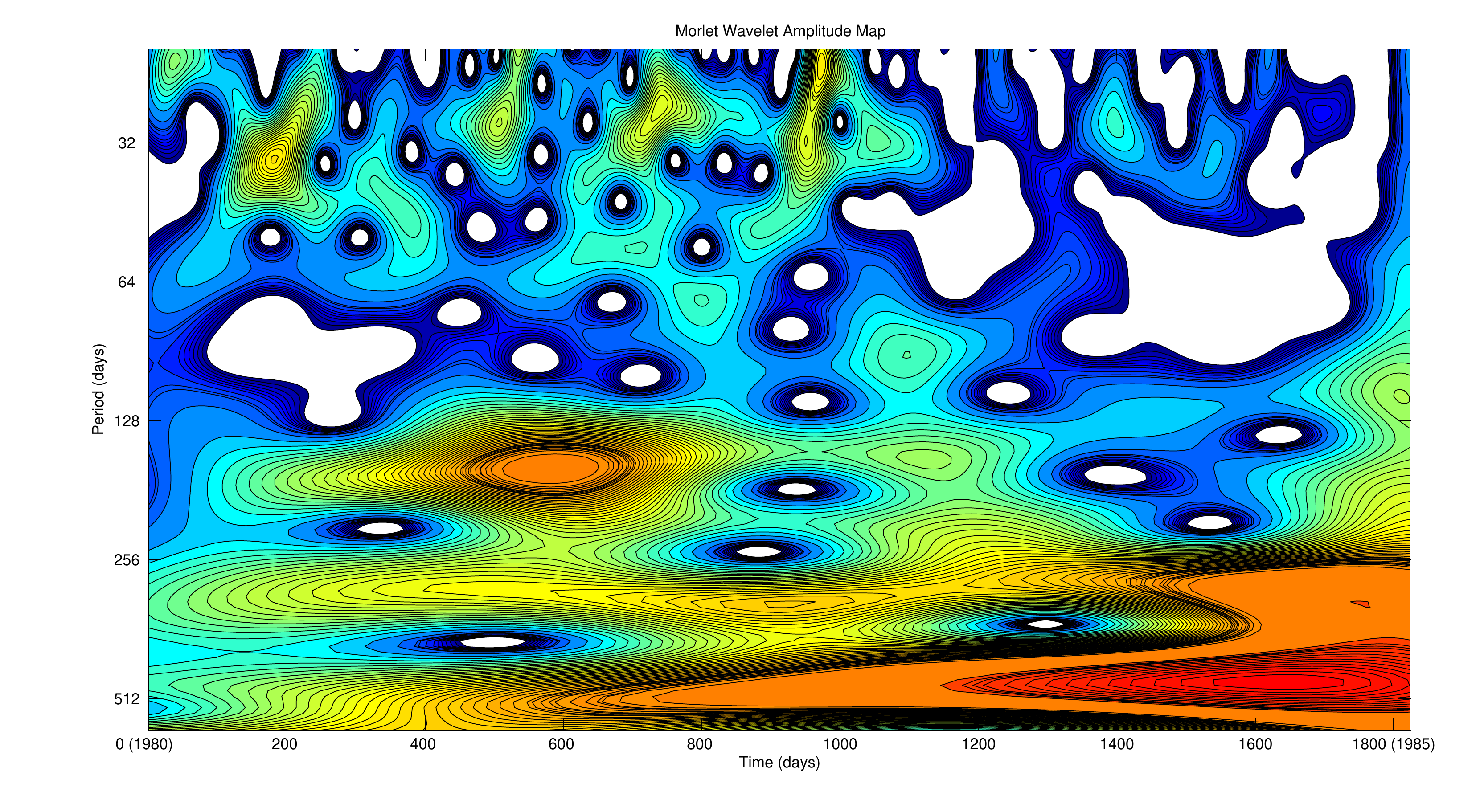}
\end{center}
\caption{Morlet Wavelet Amplitude Map of the NSO/Kitt Peak data. The local wavelet power spectrum of
the record of sunspot areas for periods for between about 22 days and 1.5
years. We also can to visual other periodocities, such as,
$\sim$30--days (diferential rotational period), 158--days, 1--year (seasonal period) and 1.3 years ($\sim$ 474.5 days) periodicities. The maximum variance corresponds to 1.3 years.}
\label{fig1}
\end{figure}

\begin{figure}[!htb]
\begin{center}
\includegraphics[width=0.85\textwidth]{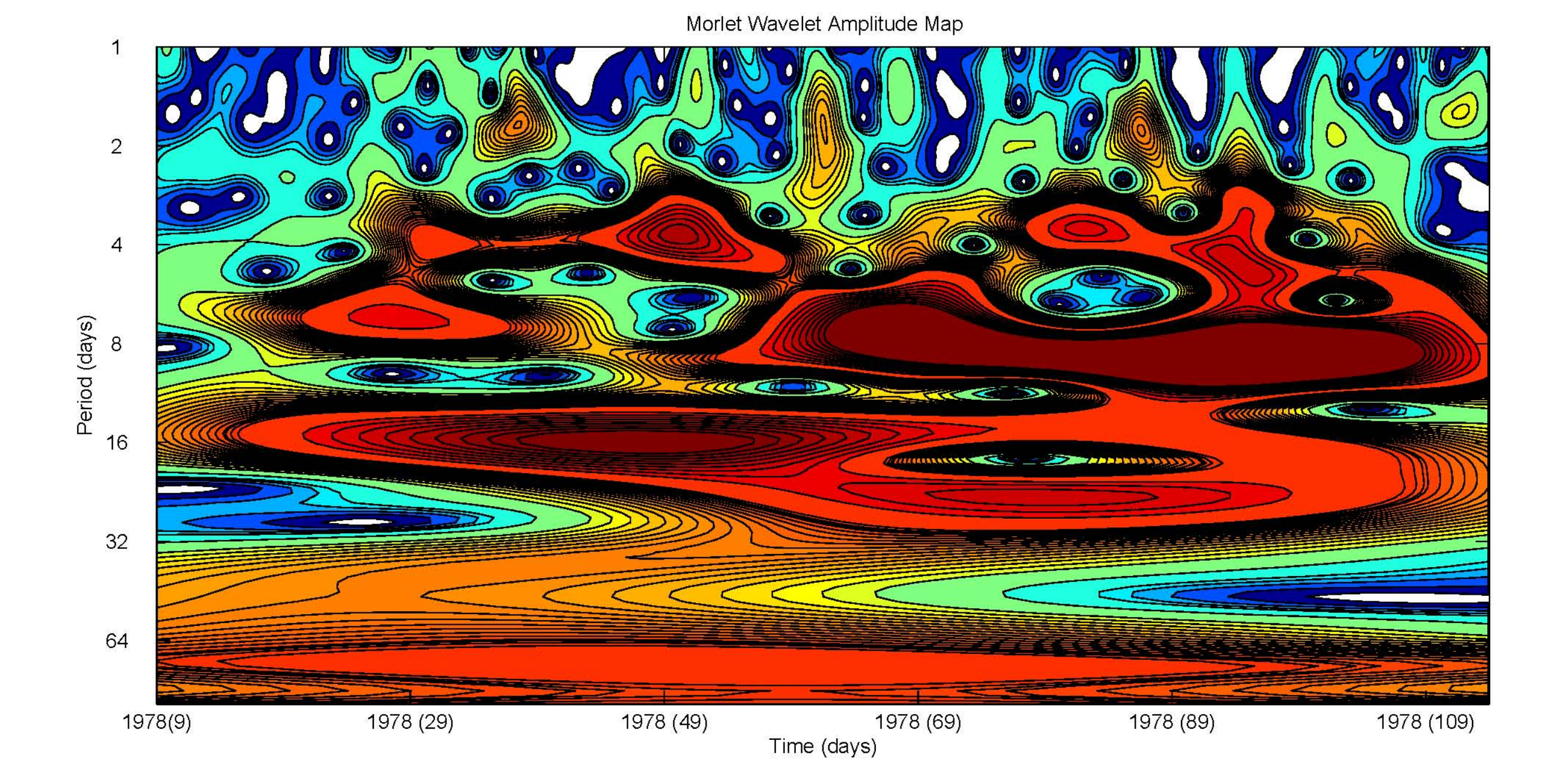}
\end{center}
\caption{Morlet Wavelet Amplitude Map of the daily averages of the magnetic field strength B versus time measured by Voyager 1 (V1) during
1978, corresponding from 2 to 3 AU. In this figure we can see that the average diferential rotation is less than of the Fig. \ref{fig1}. This effect can be caused by dispersion of solar magnetic field on the distance.}
\label{fig2}
\end{figure}

\begin{figure}[!htb]
\begin{center}
\includegraphics[width=0.85\textwidth]{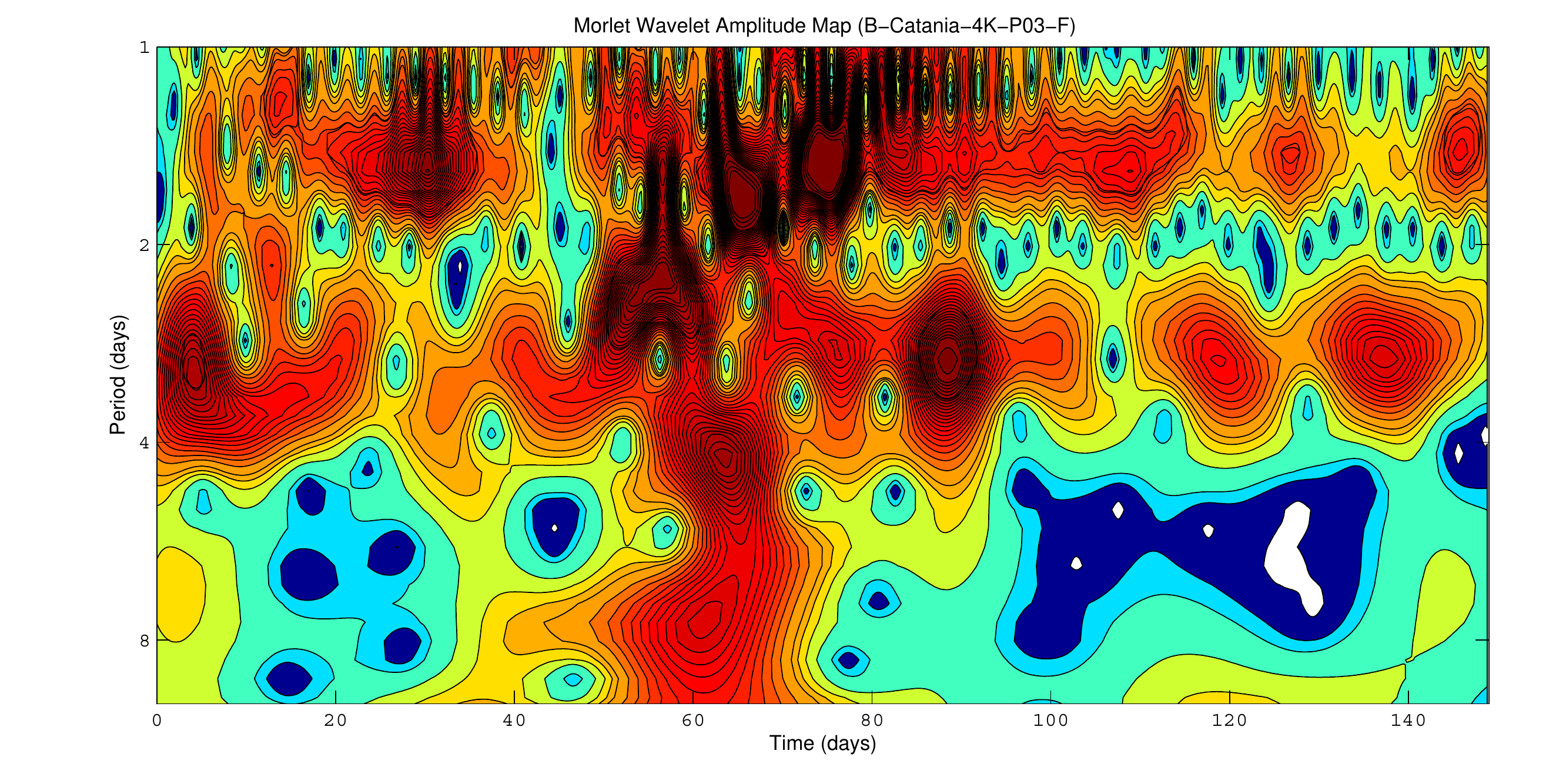}
\end{center}
\caption{Morlet Wavelet Amplitude Map of the synthesized light curve in one CoRoT passbands for a main-sequence star with T$_{eff}$ = 6000 K, log $g$ = 4.5 (cgs units), a rotation period of 3 days and a facular behaviour of type F. The period of 1.5 days corresponds to a spot 180$^{o}$ shifted longitude.}
\label{fig3}
\end{figure}

\section{Results and Conclusions}
Figures \ref{fig1}, \ref{fig2} and \ref{fig3} show the Morlet Wavelet Amplitude Maps, respectively corresponding to oscillations in the photospheric magnetic field of the Sun (NSO/Kitt Peak data), the daily averages of the magnetic field strength B versus time measured by Voyager 1 (V1) during 1978, and synthetic light curve produced by A. F. Lanza. The NSO/Kitt Peak data, as well as Voyager 1 and synthetic light curve data, are up to date the best proxies for the stellar light curves that will be obtained by CoRoT Space Mission. 

Several time series will be used as example of
wavelet analysis. These series include the NSO/Kitt Peak data used to measure
the 1.3--year (rotation of the Sun near the base of its convection zone)
and 158--day (high--energy solar flares related to a periodic emergence of magnetic flux that appears near the maxima of some solar cycles) periodicities as well as rotational period of $\sim$ 30 days\cite{has} (see Fig. \ref{fig1}). 
We have used a Wavelet analysis
(Hempelmann \& Donahue 1997; Hempelmann 2002; Lanza et al. 2003)\cite{hel,hem,lanza} to recover information on
the solar rotation rate from its {\it stellar-like} light curve. We also apply the wavelet
analysis in Sunspot data, reproducing results found in the literature
(Oliver et al. 1998; Krivova \& Solanki 2002)\cite{oliver,krivova}. We also analyze synthetic
light curve produced by a {\it theoretical simulator}, developed by the Group of Stellar Astronomy at Natal, obtaining well--defined periodicities compared with the real values.

We confirm the results obtained by analysis based
on the Lomb-Scargle periodogram and by the Phase
Dispersion Minimization method (PDM) for period search using the PERANSO
program\cite{peranso}. But, in contrast with these methods, the wavelet procedure makes
possible a global and local analysis of the periodicities\cite{strass}.

The period of rotational modulation changes during the
solar cycle because the variability of: first, the latitude of the
activity belts and
the mean lifetime of the surface features. The rotational
modulation signal can be masked by the active region evolution due to its
variability.

The use of wavelet applied in a signal, demonstrated to be a powerful
tool for analysis of signals with non--stationary features. On the other
hand, we can through the
wavelet method establish clearly the presence of a persistent signal. But,
through the decomposition at level (or frequency)\cite{poly} we can obtain other
periodicities that are not visible when we treat simultaneously
all--frequencies.

\section*{Acknowledgments}
This work was been supported by continuous grants from CNPq Brazilian Agency, by a PRONEX grant of the FAPERN Rio Grande do Norte Agency.


\end{document}